\documentclass{article}
\usepackage{graphicx} 
\usepackage{amsmath, amsfonts, amssymb}
\usepackage[colorlinks]{hyperref}
\usepackage[color=yellow]{todonotes}
\usepackage{tikz,tikz-cd}
\usepackage{parskip}
\usepackage[margin=3cm]{geometry}
\usepackage{subcaption}
\usepackage{float}
\usepackage{authblk}
\usepackage{dsfont}
\usepackage{multirow}

\hypersetup{
    colorlinks = true,
    citecolor = blue,
    linkcolor = blue,
    urlcolor = blue
}

\title{Critical latitude in global quasi-geostrophic flow on a rotating sphere}

\author[1]{Arnout D. Franken}
\author[2]{Erwin Luesink}
\author[3]{Sagy R. Ephrati}
\author[1,4]{Bernard J. Geurts}
\affil[1]{Mathematics of Multiscale Modelling and Simulation, Department of Applied Mathematics, Faculty EEMCS, University of Twente, PO Box 217, 7500 AE Enschede, The Netherlands}
\affil[2]{Korteweg-de-Vries Institute, University of Amsterdam, PO Box 94248, Science Park 107, 1090 GE Amsterdam, The Netherlands}
\affil[3]{Department of Mathematical Sciences, Chalmers University of Technology, 412 96 Gothenburg, Sweden}
\affil[4]{Multiscale Physics, Center for Computational Energy Research, Department of Applied Physics, Eindhoven University of Technology, Eindhoven, The Netherlands}

\date{\today}

\begin{document}

\maketitle

\begin{abstract}

In this paper, we study geostrophic turbulence without external forcing or dissipation, using a Casimir-preserving numerical method. The research examines the formation of large zonal jets, common in geophysical flows, especially in giant gas planets. These jets form due to the east-west stretching of vortices, influenced by the gradient of the Coriolis parameter, leading to a critical latitude beyond which jets do not form.

Using a global quasi-geostrophic model with a fully latitude-dependent Coriolis parameter, we investigate this critical latitude, which is theorized to depend only on the product of the Rossby number and the Lamb parameter. By simulating random flow fields, the critical latitude was identified through zonally averaged zonal velocity profiles.

Results align with geostrophic theory, especially near typical Rossby and Lamb parameter values for Earth's atmosphere. However, in the regime of weak rotation (high Rossby numbers) and strong stratification (high Lamb values), no clear critical latitude emerges; instead, zonal jet amplitude and width decrease gradually towards the poles. This research paves the way for further study of jet dynamics under a fully latitude-dependent Coriolis parameter.
\end{abstract}

\section{Introduction}
\label{sec:intro}

Planetary flows, encompassing both oceanic and atmospheric dynamics, exhibit highly complex behavior characterized by a wide range of nonlinearly interacting temporal and spatial scales. At large spatial scales, these dynamics are predominantly governed by the geostrophic balance, where the Coriolis force and pressure gradient force are in near equilibrium\cite{vallis2017atmospheric}. This balance is fundamental in determining large-scale circulation patterns and can be effectively described by the barotropic quasi-geostrophic (QG) equations, which approximate the evolution of potential vorticity in a shallow layer on a rotating sphere~\cite{pedlosky2013geophysical}. 

In the quasi-geostrophic regime, the flow can be characterized using the non-dimensional Rossby number and the Lamb parameter~\cite{schubert2009shallow}. The Rossby number measures the ratio of inertial to Coriolis forces, while the Lamb parameter compares buoyancy to vertical advection forces. In the shallow-water approximation, this ratio can be interpreted as the ratio between the planetary rotational speed and the internal gravity wave phase speed. 

In this work, we investigate the influence of the Rossby number ($Ro$) and the Lamb parameter ($\gamma$) on the overall dynamics, particularly focusing on the formation of circumferential planetary-scale jets under suitable flow conditions. These jets are concentrated around the equator to varying degrees, and their spatial localization in the azimuthal direction is characterized by the so-called critical latitude $\lambda$~\cite{theiss2004equatorward}. This critical latitude separates two distinct flow regimes: turbulent jet-like flows at lower latitudes and homogeneous turbulence-like flows at higher latitudes. 

Our study employs high-fidelity simulations with Casimir-preserving numerics to quantify the dependence of the critical latitude on the Rossby and Lamb parameters, extending earlier work on $\beta$-planes and $f$-planes~\cite{franken2024zeitlin}. The use of a structure-preserving integrator allows us to simulate geostrophic flow for long times in the absence of external forcing. Furthermore, the conservation of total enstrophy and higher-order moments of potential vorticity implies that no additional closure modelling is needed at the smallest resolved length scales.

In this work, we present a quantitative study of the parameter dependence of the critical latitude in global geostrophic turbulence. For Rossby numbers and Lamb parameters close to those observed in most common applications, we find clear numerical evidence of a critical latitude for jet formation, aligning closely with predicted values from theory. In regimes of weak rotation or weak stratification effects, zonal jet amplitudes only slowly attenuate near the poles, implying that no clear separation between flow regimes exists in these cases. These results can serve as a point of reference for a more nuanced understanding of geostrophic flow in the transition from wave-dominated to turbulence-dominated dynamics.

The organisation of the paper is as follows. In Section~\ref{sec:equations}, we introduce the dynamical equations on the sphere and discuss the main parameters that characterize the flow. The computational model is then presented in Section~\ref{sec:zeitlin}. Quantitative results for the critical latitude in a wide range of parameter values are presented in Section~\ref{sec:critical_lat} and concluding remarks are gathered in Section~\ref{sec:conclu}.

\section{Quasi-geostrophic model for flow on the sphere}
\label{sec:equations}

Geophysical fluid flows feature a large range of scales length and time scales, leading to enormous complexity in the multiscale dynamics due to the highly nonlinear interaction between scales. This has led to the development of the field of geophysical fluid dynamics (GFD) in which a hierarchy of simplified models have been developed for approximating the dynamics in specific flow regimes~\cite{chai2016understanding,luesink2021stochastic}. In particular, when considering sufficiently large length scales, the Coriolis force is dominant over inertial forces, leading to an approximate balance between the Coriolis force and pressure forces called the \textit{geostrophic balance}~\cite{pedlosky2013geophysical}. A first-order expansion of the equations of motion in terms of this balance leads to the quasi-geostrophic (QG) equations\cite{verkley2009balanced}.

The QG equations describe the motion of a fluid on a sphere of radius $R$, rotating at an angular velocity $\Omega$. The fluid occupies a shallow dynamic upper layer of average thickness $H$ on top of a fixed topography, which we assume to be flat in the remainder. The Coriolis parameter is given by $f = 2 \Omega \sin \phi$, where $\phi$ is the latitude on the sphere such that the equator is given by $\phi = 0$.

Typically, the QG equations are studied in the so-called $\beta$-plane approximation~\cite{vallis2019essentials}. In that case, the fluid domain is a Cartesian tangent plane to the sphere at a certain reference latitude on which the Coriolis parameter is linearized as $f_0+\beta y$, where $y$ is the meridional coordinate. Taking $\beta=0$, the classic $f$-plane approximation is recovered, in which no zonal jets are formed due to the absence of a gradient in the Coriolis parameter~\cite{rhines1975waves}. The typical eddy radius is given by the Rossby deformation radius $L_D= \sqrt{g'H}/|f_0|$, which is constant on the tangent plane. On the $\beta$-plane, zonal jets can be generated by Rossby waves, which has made it instrumental in the past for understanding the formation of these jets. However, \cite{theiss2004equatorward} has pointed out that this approach fails to account for the qualitatively different behaviour of the flow near the equator compared to the polar regions. This is because, on the sphere, the Rossby deformation radius $L_D$ is not constant due to the variation of the Coriolis parameter, which modifies the ability of Rossby waves to induce a zonal flow. In particular, at increasing latitudes, the Coriolis parameter is larger, leading to a smaller Rossby radius, which at sufficiently high latitudes even eliminates the ability of Rossby waves to elongate eddies into zonal structures. 

To include latitude-dependent Rossby-wave effects in the model, \cite{theiss2004equatorward} proposes to include the fully latitude-dependent Coriolis parameter in tangent plane simulations and shows that this model correctly predicts equatorward kinetic energy transfer as well as the emergence of a critical latitude in the dynamics. This critical latitude marks the transition between the equatorial region, where zonal jets form, and the poleward region, characterized by largely isotropic turbulence, effectively separating these distinct flow regimes.

The critical latitude characteristic has been studied numerically for situations with a very low critical latitude. This is motivated by the fact that the gradient of the Coriolis parameter is largest at low latitudes, thus giving rise to a sharper transition in the dynamics. For many natural situations, such as the winds on Jupiter, zonal jets form even at high latitudes, where the spherical geometry of the domain starts to play an important role. In this work, we therefore consider the quasi-geostrophic equations on a full sphere, eliminating the need for additional approximations as seen in literature until now. Studying fluid dynamics on the sphere allows for great flexibility in the choice of parameters since all latitudes are included in the domain. Furthermore, working on the sphere removes the need for artificial meridional and zonal boundary conditions that might impact the dynamics in an unphysical way.

The formulation of quasigeostrophic equations on a spherical geometry was independently rediscovered by \cite{schubert2009shallow} and \cite{verkley2009balanced}, based on earlier work by \cite{kuo1959finite} and \cite{charney1962stability}. Recently, a geometric derivation of these equations was presented in \cite{luesink2024geometric}, highlighting the Hamiltonian structure of the equations, as well as identifying the key conserved quantities. Based on this, a computational model was constructed in \cite{franken2024zeitlin} using recent work presented in~\cite{cifani2022casimir,cifani2023efficient}, which enables simulations of quasi-geostrophic turbulence on a sphere while conserving all resolved moments of potential vorticity, in addition to energy. 

The single prognostic variable in the QG equations is the potential vorticity (PV), which is conserved along fluid trajectories. The fluid velocity can be calculated from the associated stream function $\psi$ which can in turn be diagnosed from the PV. Thus, the dynamics are given by the following equation
\begin{equation}
    \frac{\partial q}{\partial t} + \mathbf{u} \cdot \nabla q = 0,
\end{equation}
where $q$ is the potential vorticity and $\mathbf{u}$ the horizontal velocity field. Since the horizontal velocity field is divergence-free, there is an associated stream function $\psi$ such that
\begin{equation}
    \mathbf{u} = \nabla^\perp \psi,
\end{equation}
where we adopt the notion from \cite{luesink2024geometric} to denote the right-handed rotation of a vector in a two-dimensional setting by 90 degrees with $(\cdot)^\perp$. This is equivalent to taking the three-dimensional curl with the outward normal $\hat{\mathbf{r}}\times \cdot$ and then projecting on the horizontal dimensions. Using the stream function, the evolution of potential vorticity can be rewritten as
\begin{equation}
    \frac{\partial q}{\partial t} + \left\{\psi,q\right\} = 0
    \label{eq:pvtransport}
\end{equation}
where the bilinear operation $\{a,b\} = \nabla^\perp a \times  \nabla b$ is the canonical Poisson bracket, which is sometimes referred to as the Jacobian in literature (\cite{theiss2004equatorward}). The diagnostic equation for the stream function on the sphere is given by \cite{franken2024zeitlin}
\begin{equation}
    q = \nabla^2 \psi + f - L_D^{-2}\psi
    \label{eq:pvdiagnostic}
\end{equation}
where $L_D$ is the barotropic Rossby deformation radius, which in this case is fully latitude-dependent and given by
\begin{equation}
    L_D = \frac{\sqrt{gH}}{|f|}
    \label{eq:RD}
\end{equation}
Note that at the equator, $f=0$, resulting locally in an unbounded deformation radius. However, this does not lead to unbounded solutions, since $L_D^{-2}$ in equation \ref{eq:pvdiagnostic} is always well-defined and bounded. This formulation is similar to the model used in \cite{theiss2004equatorward}, with the important difference that the present model is formulated on a spherical domain, which changes the coordinate expression of the Poisson bracket in (\ref{eq:pvtransport}). For more details on the geometrical aspects of the equations, we refer the reader to \cite{luesink2024geometric}.

For the following analysis, it is convenient to write the equation in non-dimensional form. We rescale the horizontal distances using the radius of the sphere $R$, and denote with $V$ the characteristic fluid velocity scale, which induces a natural time scale of $T = R/V$. Using this scaling, we can write the equations in non-dimensional form as
\begin{align}\label{eq:QG}
    &\frac{\partial q}{\partial t} + \left\{\psi,q\right\} = 0 \\
    &q = \nabla^2 \psi + \frac{2\mu}{\mbox{Ro}} - \gamma \mu^2 \psi
\end{align}
where $\mu = \sin \phi$ and $\mbox{Ro}=V/(\Omega R)$ is the global Rossby number. Furthermore, the gravitational effects are scaled by the Lamb parameter $\gamma$ given by
\begin{equation}
    \gamma = 4\frac{R^2\Omega^2}{gH}.
\end{equation}
The QG model can describe the formation of solutions in the form of a number of jets in the circumferential direction. The structure of these solutions can be modified by external rotation of the sphere. In Figure~\ref{fig:qg_nice} we present a snapshot displaying qualitative features of the QG solution for the zonal component of the horizontal velocity field. Large-scale structure is observed next to much finer-scaled complex flow, loosely referred to as QG turbulence.

To simulate the flow governed by QG on the sphere, we adopt a particular structure-preserving discretization to which we turn next.

\section{Zeitlin truncation as structure-preserving discretization}
\label{sec:zeitlin}

In this section, we detail the numerical method adopted for the discretization of the QG equations on the sphere. We utilize the numerical framework that was recently developed for the quasi-geostrophic equations on the sphere in \cite{franken2024zeitlin}. By discretizing the Poisson structure of the equation, this approach leads to a fully energy- and enstrophy-conserving approach.

A well-known property of incompressible fluids in two dimensions is the conservation of any integrated function of vorticity~\cite{zeitlin2018geophysical}, as well as an energy, given in this case by 
\begin{equation}
    \mathcal{H} = \frac{1}{2}\int_{S^2} \psi (q-2\mu) \, \mbox{d}x
\end{equation}
The numerical method is based on a truncated expansion of functions $u\in \mathcal{C}^\infty(\mathcal{S}^2)$ on a unit sphere in terms of spherical harmonics $\{Y_{lm}\}$ with $l$ the degree and $m$ the order, given as
\begin{equation}
    u(\mathbf{x}) \approx \sum_{l=0}^{N-1}\sum_{m=-l}^l u_{lm} Y_{lm},
\end{equation}
where $N$ is the adopted resolution. In order for the evolution of the spherical harmonics coefficients $q_{lm}$ of potential vorticity to be conservative, a mapping $\Pi_N : C^\infty(S^2) \to \mathfrak{su}(N)$ is used to represent functions as complex skew-symmetric $(N\times N)$ matrices. For a function $u$ on the sphere, this projection is given as
\begin{equation}
    U:= \Pi_N(\mathcal{F}) = \sum_{l=0}^{N-1}\sum_{m=-l}^l u_{lm}T_{l,m}^N,
\end{equation}
where $T_{lm}^N\in \mathds{C}^{N\times N}$ are the basis skew-symmetric \textit{matrix} elements~\cite{zeitlin2004self}. Analogous to the continuum case, these basis elements are defined as the eigenvectors of the \textit{discrete Laplacian} $\Delta_N$ such that
\begin{equation}
    \Delta_N T_{lm}^N = l(l+1) T_{lm}^N,
\end{equation}
where the discrete Laplacian $\Delta_N$ is given in \cite{hoppe1998some}. The main property of this projection approach is that potential vorticity transport is now approximated by a unitary transformation of the potential vorticity matrix. This is given by the property of the projection operator
\begin{equation}
    \Pi_N \{ \psi, q \} = \frac{N^{3/2}}{16\pi}[ \Pi_N \psi, \Pi_N q ] + \mathcal{O}(N^{-2}),
\end{equation}
which implies that the time evolution of potential vorticity is well approximated by the matrix commutator $[A,B]=AB-BA$.
For a more detailed overview of this projection method applied to the 2D Euler equations, we refer the reader to \cite{modin2020casimir} and for a more general overview to \cite{modin2024two}.

Given this matrix representation of the transport equation (\ref{eq:QG}), time integration can be performed using an isospectral Lie-Poisson time integrator as described in \cite{franken2024zeitlin}. This ensures the conservation of energy, as well as all integrated moments of potential vorticity up to power $N-1$. This is a crucial feature for retaining long-time characteristics of the dynamics.

\begin{figure}
    \centering
    \includegraphics[width=.8\columnwidth]{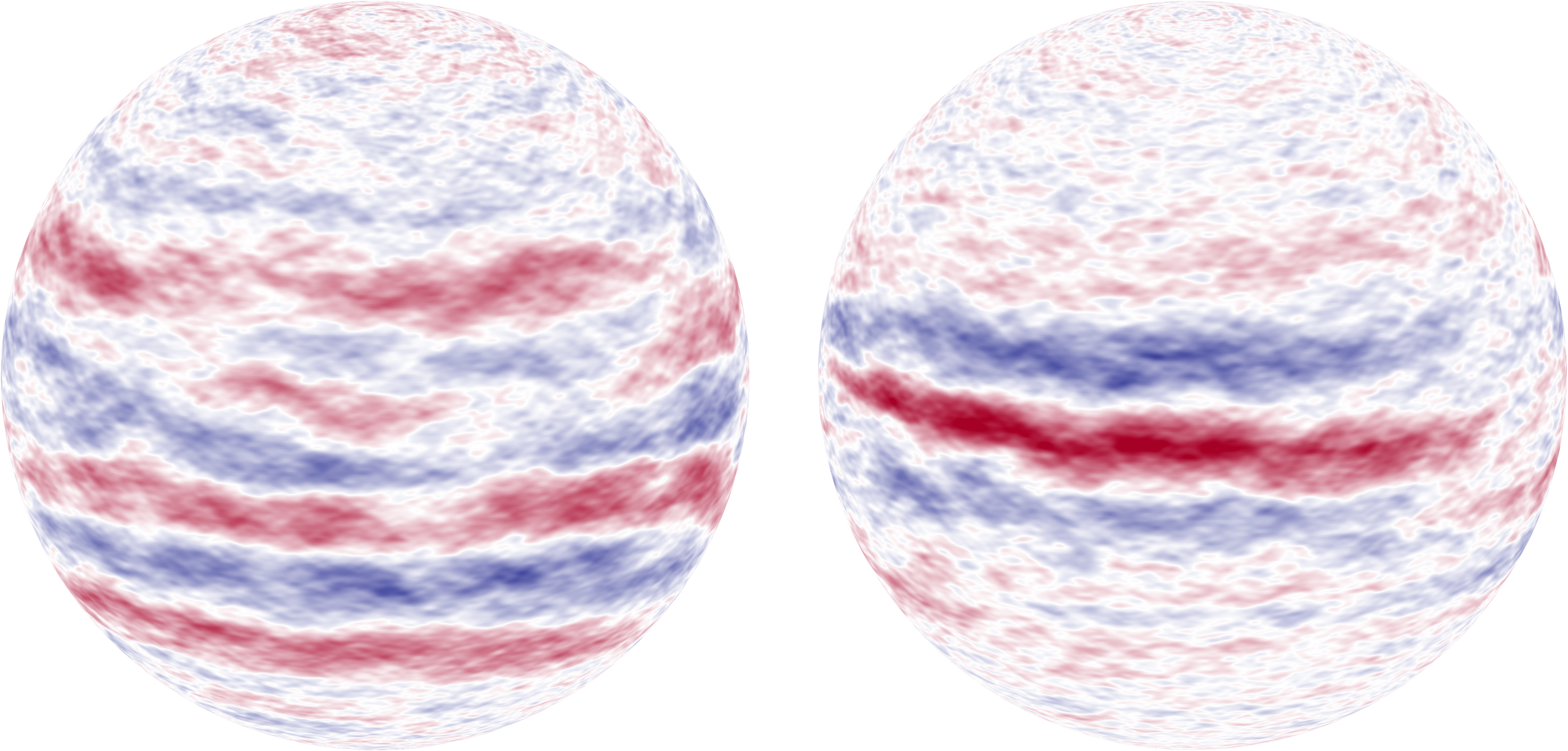}
    \caption{Snapshot at $T= 8\,000$ days of the zonal velocity on the sphere with $\mbox{Ro} = 1/250$ for $\gamma=500$ (left) and $\gamma = 2000$ (right). Red and blue colours indicate prograde and retrograde motion respectively.}
    \label{fig:qg_nice}
\end{figure}

The global dynamics and how this depends qualitatively on the Lamb parameter is illustrated in Figure~\ref{fig:qg_nice}. This figure shows a snapshot of the zonal component of the velocity field after 8000 days for simulations at a Lamb parameter value of 500 (left) and 2000 (right), evolving from a random initial condition. These results were obtained at a resolution of $N=512$, containing all spherical harmonics modes up to degree $l=511$. The velocity snapshots show the main feature of global geophysical fluid dynamics, which is the emergence of zonal jets. These are elongated structures in the east-west direction that form the main flow feature at large scales. Crucially, the formation of these jets seems to be strongly dependent on the value of the Lamb parameter, as the snapshot on the right only shows one eastward-moving jet and two westward-moving jets, while the snapshot on the left shows jet development at much higher latitudes. In particular, the snapshot on the right indicates the emergence of a critical latitude beyond which no jets can form as was studied earlier in~\cite{theiss2004equatorward}. In the following section, the dependence of this critical latitude on the Lamb parameter as well as the Rossby number will be studied numerically in detail using simulations at various parameter values.

\section{Critical latitude}
\label{sec:critical_lat}

In this section, we first introduce the critical latitude and the way this quantity is computed from the simulations in Subsection~\ref{subsec:latdef}. The general features of the jets that emerge from the simulations are presented in Subsection~\ref{subsec:resolved} in which we also describe the spectrum associated with the dynamical interactions. Finally, in Subsection~\ref{subsec:critlat} the dependence of the critical latitude on the Rossby number and Lamb parameter are presented.

\subsection{Definition of critical latitude}
\label{subsec:latdef}

In geostrophic turbulence, two length scales are known to play a major role in the characterization of the flow. The first of these, the Rossby deformation length, has been introduced in (\ref{eq:RD}). The second important length scale is the Rhines length, named after Peter Rhines (\cite{rhines1975waves,rhines1979geostrophic}), who first developed a quantitative theory of the interplay between waves and turbulence in geostrophic flows. The Rhines length is given by
\begin{equation}\label{eq:rhinesscale}
    L_R = \sqrt{V/\beta}
\end{equation}
Here, $V$ is the chosen reference velocity and $\beta$ refers to the gradient of the Coriolis parameter. Similar to the Rossby deformation length, the Rhines length takes a constant value in a $\beta$-plane model. On the sphere, however, both length scales vary significantly with latitude. The Rossby length scale is inversely proportional to the value of the Coriolis parameter which takes its maximum value at the poles while being zero at the equator of the sphere. The dependence of the Rhines length on latitude is expressed by the gradient of the Coriolis parameter, given by $\beta = 2\frac{\Omega}{R} \cos \phi$ in which $\Omega$ is the angular velocity of the planet with radius $R$ and $\phi$ is the latitude.

These Rossby and Rhines length scales represent two important phenomena in geostrophic flows. The Rossby length scale enters the dynamical equations as a vortex-stretching term~\cite{pedlosky2013geophysical}. It introduces a latitude-dependent gradient in the potential vorticity field on the sphere, which leads to an elongation of vortices in the east-west direction. This effect is most noticeable at large length scales where inertial forces are small compared to pressure forces and the Coriolis force. In these circumstances, the flow is dominated by Rossby waves, which are the wave solutions of the linearized equations~\cite{schubert2009shallow}. Therefore, the behaviour of geostrophic flow at scales larger than the Rossby scale is often characterized as wave-like~\cite{salmon1982geostrophic}. On the other hand, the Rhines scale (\ref{eq:rhinesscale}) represents the turbulent length scale, also called the \textit{arrest scale}~\cite{penny2010suppression}. The presence of a gradient in the Coriolis parameter leads to isotropy in the inverse energy cascade that is typical for two-dimensional flows~\cite{kraichnan1967inertial}. In the absence of rotation, energy cascades to scales that minimize the total wave number $\mathbf{k}$, thus cascading to the largest length scales available. In the presence of rotation, however, energy cascades to scales that minimize $\mathbf{k}^2 + L_D^{-2}$~\cite{salmon1998lectures}. In other words, the energy cascade is arrested at a particular length scale $L_D$, implying that most energy is contained in these scales. Additionally, this cascade is highly anisotropic due to the latitude-dependence of $L_D$ in a global domain. 

In \cite{theiss2004equatorward}, the critical latitude is defined using the ratio 
\begin{equation}\label{eq:nonlinparam}
    r = \frac{L_D}{L_R}.
\end{equation}
This ratio is also called the nonlinearity parameter for QG turbulence, as it measures the relative strength of the nonlinear advection term compared to hydrostatic and Coriolis effects~\cite{klocker2016regime}. It is argued in \cite{theiss2004equatorward} that Rossby waves can only induce alternating zonal flow if $r>1$. This is based on the analyses presented in \cite{rhines1975waves, rhines1979geostrophic}. A similar observation was made in~\cite{klocker2016regime}, who identified the regions where $r<1$ with isotropic turbulence (confusingly, the inverse of the nonlinearity ratio was used). Since the Rossby deformation length $L_D$ is zero at the poles and becomes infinitely large at the equator, the line $r=1$ represents the critical latitude above which zonal jets cannot be formed. In many planetary systems, including the Earth's atmosphere, this leads to the well-known stable global circulations near the equator, with more homogeneous weather patterns at higher latitudes~\cite{vallis2017atmospheric}. 

Using the model equations, we find a theoretical estimate of the critical latitude $\phi_c$ from
\begin{equation}
    \frac{L_D}{L_R} = \sqrt{\frac{g'H\beta}{f^2U}} = \sqrt{\frac{g'H \cos \phi_c}{2\Omega RU\sin^2\phi_c}} =1
\end{equation}
After some algebraic manipulation, the critical latitude $\phi_c$ can be determined from inverting
\begin{equation}
    \cos \phi_c = \frac{\sqrt{1+s^2} - 1}{s}, \quad \mbox{where} \quad s = \frac{U}{R} \frac{\gamma}{\Omega} = \mbox{Ro} \, \gamma
    \label{eq:CL}
\end{equation}
This is similar to the expression found in \cite{theiss2004equatorward}. The result shows that the critical latitude is only dependent on a single parameter $s$, which represents the ratio between the strength of hydrostatic effects compared to the Coriolis effect. 

Equation (\ref{eq:CL}) reveals that the critical latitude is only dependent on the parameter $s$, which represents the ratio between gravitational and rotational effects as seen from the ratio $\gamma/\Omega$. Furthermore, the critical latitude increases monotonically with this ratio. When $\Omega$ increases (leading to a lower Rossby number), the relative influence of the Coriolis parameter increases, which extends the domain in which eddies are elongated in the east-west direction such that they form zonal jets. On the other hand, an increasing $\gamma$ implies that eddies are prevented from coalescing to larger scales, thus weakening the ability of the beta effect to stretch eddies zonally~\cite{klocker2016regime}. 

In the next subsection, we provide the details regarding numerical simulations of geostrophic turbulence on the sphere. This allows us to give a phenomenological description of the critical latitude leading to the numerical procedure for determining its value. These numerical results will be compared to the theoretical results presented in this subsection.

\subsection{Resolved dynamics of zonal jets}
\label{subsec:resolved}

We study the emergence of zonal jets in various flow settings by simulation of barotropic QG turbulence at several values of the Rossby number and the Lamb parameter. Using the numerical method described in the previous section, we select several different rotation velocities for the planet leading to different Rossby numbers, as well as different shallow water configurations leading to different values for the Lamb parameter. For both parameters, we select values in the realistic range for well-known planetary flows. Typical values for the global Rossby number range from $10^{-3}$ for oceanic flow to $10^{-1}$ for atmospheric flow~\cite{luesink2024geometric}. The Lamb parameter is in many applications mainly influenced by the height of the fluid column, with a reported range between $10^1-10^4$ for most common situations~\cite{schubert2009shallow}.

Similar to the work of \cite{theiss2004equatorward}, we simulate flow on a planet similar to the Earth, giving a radius of the planet of $R=6\cdot 10^6$ m. However, the angular velocity of the planet is varied in order to study the influence of the Rossby number. Multiple simulations are performed at different values of the global Rossby number. Since the initial condition is tuned such that the maximum jet velocity is of order unity for all simulations, the global Rossby number can be set by choosing an appropriate value for the angular velocity $\Omega$. We select Rossby numbers in the geostrophic regime, $\mbox{Ro} = 1/125$, $1/250$, $1/500$, $1/1000$ and $1/2000$. This is done by selecting $\Omega = 2.1 \cdot 10^{-5}$, $4.2\cdot 10^{-5}$, $8.3\cdot 10^{-5}$, $1.7\cdot 10^{-4}$ and $3.3\cdot 10^{-4}$. Similarly, we select the Lamb parameter as $\gamma = 500$, $1000$, $2000$, $4000$ and $8000$. In total, these parameter choices led to 25 simulations that have been performed.

We select a spatial resolution of $N=512$ based on earlier work in \cite{franken2024zeitlin} and \cite{cifani2022casimir}. This resolution implies an angular resolution of approximately $0.5$ degrees on the sphere, which ensures that the main dynamical structures are well-captured. The flow is given a random small-scale initial PV field by setting a random phase for all modes with $40<l<60$, with a random amplitude around a reference value such that the resulting velocity field is of order unity. Starting from this initial condition, we integrate the solution in time using a time step of $\Delta t = 600$ s. We simulate for a total simulation time of $T=1.5\cdot 10^9$ seconds. This corresponds to a different number of 'days' of the planet depending on the angular velocity. Since the velocity scale of the fluid is the same for each simulation, during this time, the jets circumnavigate the planet approximately 30 times for each simulation. 

For each simulation, a theoretical estimate for the critical latitude is given by equation \ref{eq:CL}. The critical latitude marks the transition between two flow regimes; the zonal jets in the equatorial region and the more isotropic turbulent region near the poles. In \cite{theiss2004equatorward}, the critical latitude is determined from numerical experiments using the mean zonal velocity profile. These are zonal averages of the zonal component of the velocity field as a function of latitude. Additionally, the average eddy velocity is calculated as the mean absolute deviation of the velocity field from the mean zonal velocity profile. Their results show a clear transition at the predicted critical latitude. While the eddy velocity is similar between the two flow regimes, the mean zonal velocity shows a clear transition from alternating jets below the critical latitude to a zero mean velocity above the critical latitude (in the northern hemisphere).

In global simulations of freely decaying turbulence, however, the zonally averaged velocity field is not a good indicator for the presence of jets, since these jets are typically not fixed at a certain latitude. Since they do not form a stationary full band around the planet, a latitudinal circle may intersect multiple jet structures, leading to a zero average velocity. This effect is visible in Figure \ref{fig:qg_nice}, showing that jets are not fixed at a certain latitude. This is in contrast to forced turbulence, which leads to stable circumnavigating jets~\cite{scott2012structure}.

A clear indicator of the presence of zonal jets is the zonally averaged amplitude of the zonal velocity field. Since the zonal jet regime is characterized by large coherent structures, the amplitude of the zonal component of the velocity field is a measure of the amplitude of the jet. Given a decomposition of the velocity field $\mathbf{u}=(u_\phi,u_\theta)$ into its meridional and zonal components respectively, we can thus analyze the squared amplitude of the average zonal velocity $E_{zon}$ defined as
\begin{equation}
    E_{zon}(\phi) = \int_{\theta = 0}^{2\pi} u^2_{\theta}(\phi,\theta)\, d\theta.
\end{equation}
to determine the critical latitude also directly from simulations.

In the isotropic turbulent region, no preference between the zonal and meridional direction is present, leading to a uniform value of $E_{zon}$ in this regime. In the equatorial region, the presence of zonal jets leads to the velocity field being mostly aligned with latitudinal lines, leading to an oscillating profile of $E_{zon}$ with a maximum at the centre of each jet. The transition between these regimes can thus numerically be determined as the latitude at which the squared amplitude of the zonal velocity exceeds a minimum threshold. Using data from Figure 7 in \cite{theiss2004equatorward}, we can determine that the typical eddy velocity in the isotropic regime is around $0.3$ m/s. Thus, by setting the threshold for the critical latitude at $E_{zon} = 0.1$, the critical latitude can be determined for global geostrophic turbulence simulations.

\begin{figure}
    \centering
    \includegraphics[width = \columnwidth]{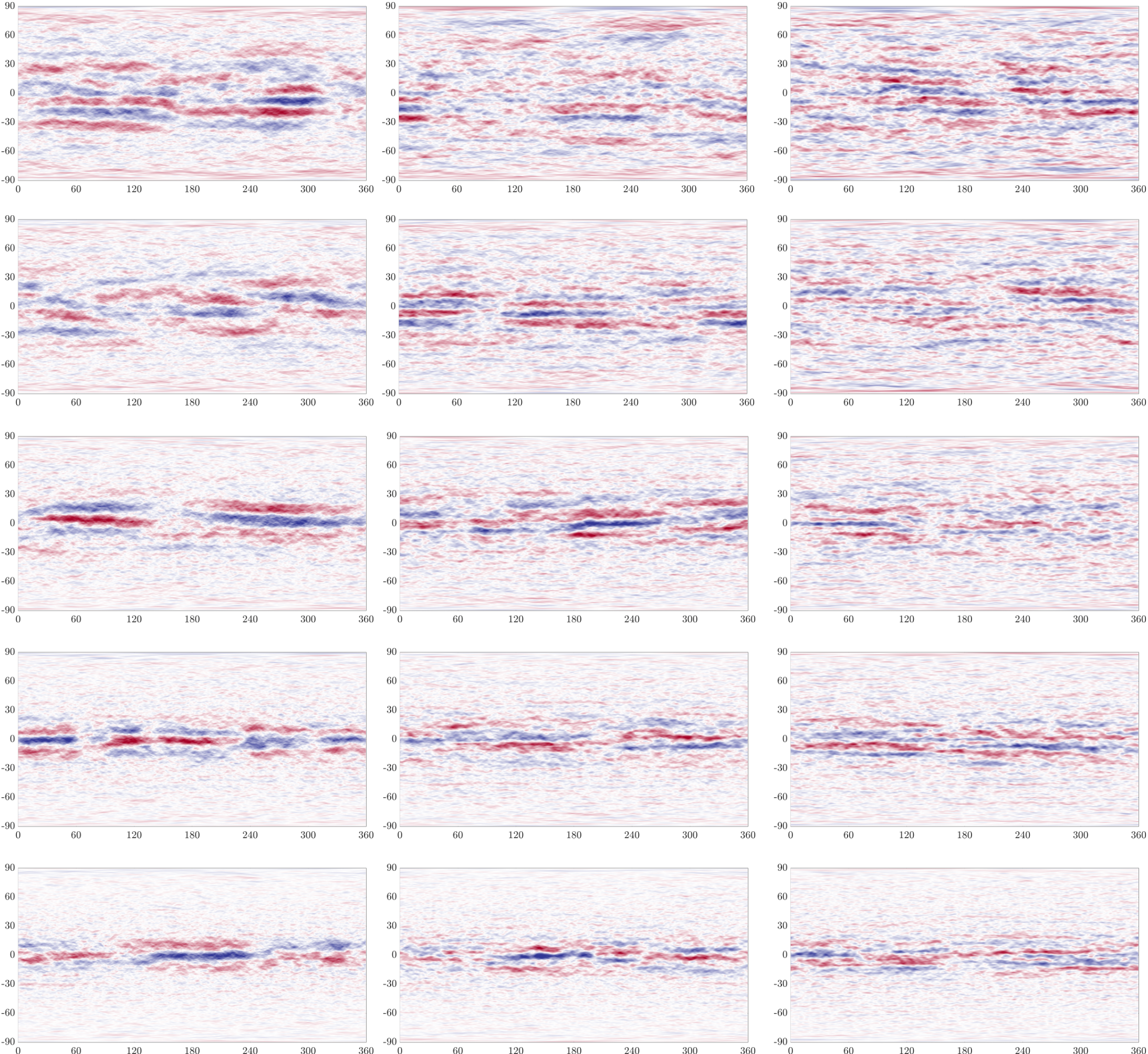}
    \caption{Snapshots of the zonal velocity fields for simulation at several different values of $\mbox{Ro}$ and $\gamma$, projected onto a latitude-longitude grid. Snapshots are taken in the well-developed regime, at the time when the jets have circumnavigated the sphere around 30 times. Red colour indicates prograde motion, while blue color indicates retrograde motion. From left to right, the figure shows simulations at $\mbox{Ro} = 1/250$, $1/500$ and $1/1000$ respectively. From top to bottom values of $\gamma = 500$, 1000, 2000, 4000 and 8000 are shown. }
    \label{fig:Vzon_table}
\end{figure}

In Figure \ref{fig:Vzon_table}, snapshots of the zonal velocity field are visualized for several simulations. These snapshots are taken after $t=1.5\cdot 10^9$ s. From left to right, a decreasing Rossby number was set, implying a faster rotation of the planet. From top to bottom the effects of an increase in the Lamb parameter are shown, indicating the result of a stronger stratification. In all simulations, zonal jets develop in the equatorial regions. These consist of elongated structures of zonal velocity, appearing in an alternating pattern across latitudes. This is consistent with the findings of \cite{franken2024zeitlin} for forced turbulence. As indicated, these jets do not fully envelop the planet but typically show a longitudinal extent of approximately 180 degrees. In all cases except those with strong rotation and weak stratification (top right), a clear transition is visible in the zonal velocity field, indicating a sharp transition between the two flow regimes. In the remaining case, jets seem to form at all latitudes, and no critical latitude is visible between the equator and the poles.

\subsection{Critical latitude as a function of Rossby and Lamb parameters}
\label{subsec:critlat}

In this subsection, we present how the simulation results for the critical latitude depend on the Rossby and Lamb parameters and compare the simulated findings with the approximate theoretical prediction for $\phi_c$.

\begin{figure}
    \centering
    \includegraphics[width = .8\columnwidth]{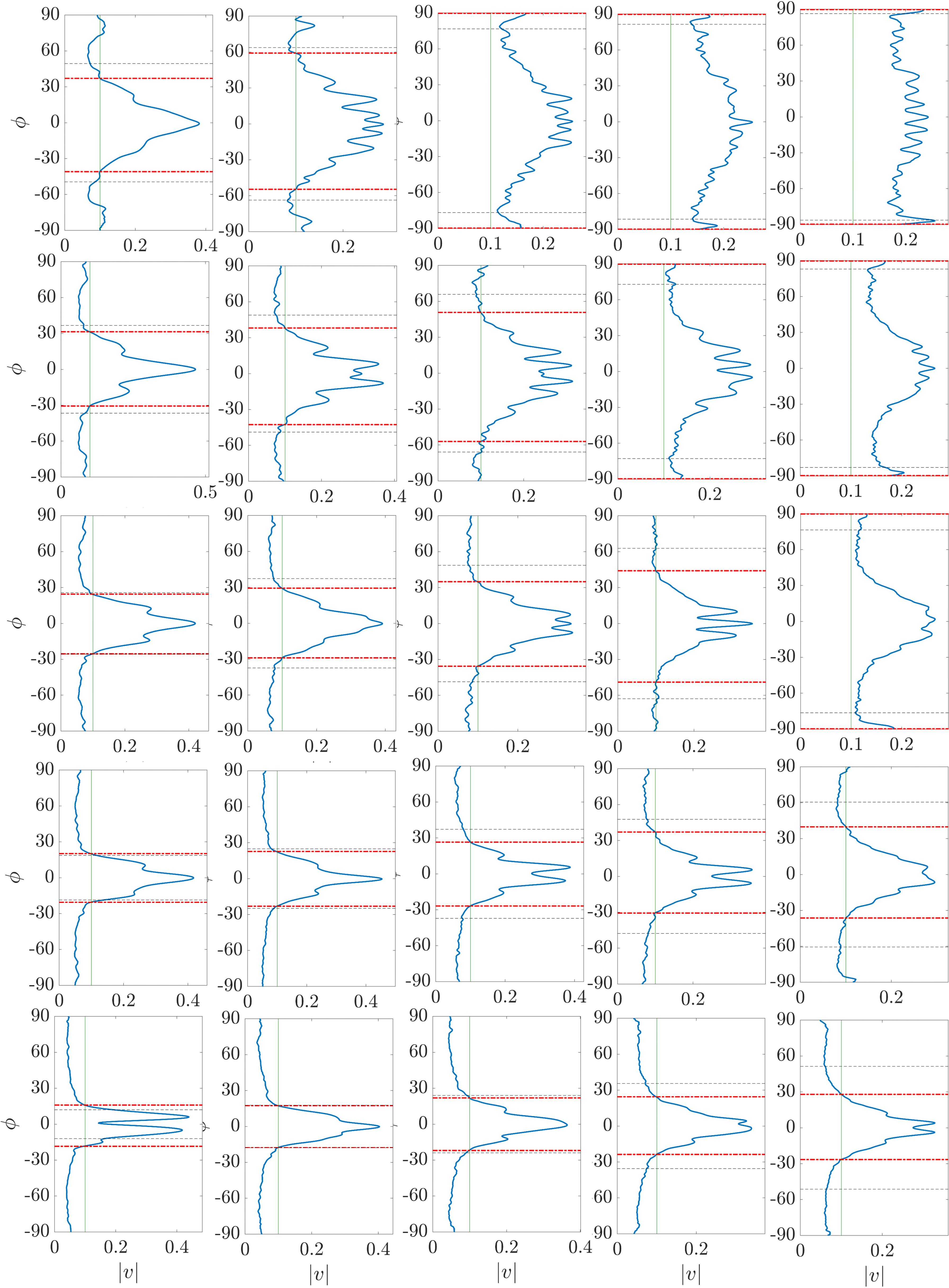}
    \caption{Average zonal velocity profiles corresponding to the simulations in figure \ref{fig:Vzon_table}, using a time average of 1000 days. From left to right, the figure shows simulations at $\mbox{Ro} = 1/125$, $1/250$, $1/500$, $1/1000$ and $1/2000$ respectively. From top to bottom values of $\gamma = 500$, 1000, 2000, 4000 and 8000 are shown. The critical latitude is given in the red dash-dotted line as the latitude at which the intensity of the absolute zonal velocity has dropped to below $|u|=0.1$ m/s. The dashed black line shows the critical latitude as calculated from equation \ref{eq:CL} with parameter $s$ computed according to the simulation parameters collected in Table~\ref{tab:CLtable}. In some cases, zonal jets form at every latitude, and the corresponding average zonal velocity does not drop below the threshold, which marks the absence of a critical latitude in those cases.}
    \label{fig:CL_table}
\end{figure}

Figure \ref{fig:CL_table} shows the average amplitude of the zonal velocity field $E_{zon}$ as a function of latitude for all simulations. A time averaging window of $1.5\cdot 10^8$ was used when calculating the profiles, corresponding to approximately 3 circumnavigations of the jets to ensure a robust statistical representation of the zonal velocity dynamics. The threshold of $E_{zon} = 0.1$ is displayed using a vertical line. As expected, in the polar regions, the flow is largely isotropic, leading to a uniform value of $E_{zon}$, while an oscillating pattern is present in the equatorial region due to the strong jets. Using the threshold value, a critical latitude was determined in both hemispheres using $E_{zon}=0.1$ m$^2$/s$^2$, which is displayed using two red horizontal lines for each simulation. The theoretical estimate of the critical latitude is shown as a dashed black horizontal line. In six cases, no critical latitude was identified from the $E_{zon}$ profiles, since it never crosses the threshold. A clear critical latitude is also visibly absent in the zonal velocity profiles in Figure~\ref{fig:Vzon_table}, showing jet formation even in the polar regions. In these cases, the critical latitudes are not defined and have not been included.

\begin{table}
\centering
\begin{tabular}{ll|lllll|}
\cline{3-7}
 &   &  \multicolumn{5}{c|}{Ro} \\ 
\cline{3-7} 
 &   &
  \multicolumn{1}{c|}{1/125} &
  \multicolumn{1}{c|}{1/250} &
  \multicolumn{1}{c|}{1/500} &
  \multicolumn{1}{c|}{1/1000} &
  1/2000 \\ \hline
\multicolumn{1}{|c|}{\multirow{5}{*}[-8ex]{$\gamma$}} &
  500 &
  \multicolumn{1}{c|}{\begin{tabular}[c]{@{}c@{}}$39.0\pm1.9$\\ $(49.5)$\end{tabular}} &
  \multicolumn{1}{c|}{\begin{tabular}[c]{@{}c@{}}$56.9\pm2.7$\\ $(63.7)$\end{tabular}} &
  \multicolumn{1}{c|}{\begin{tabular}[c]{@{}c@{}}-\\ $(76.7)$\end{tabular}} &
  \multicolumn{1}{c|}{\begin{tabular}[c]{@{}c@{}}-\\ $(81.4)$\end{tabular}} &
  \multicolumn{1}{c|}{\begin{tabular}[c]{@{}c@{}}-\\ $(86.3)$\end{tabular}} \\ \cline{2-7} 
\multicolumn{1}{|c|}{} &
  1000 &
  \multicolumn{1}{c|}{\begin{tabular}[c]{@{}c@{}}$31.1\pm0.3$\\ $(36.7)$\end{tabular}} &
  \multicolumn{1}{c|}{\begin{tabular}[c]{@{}c@{}}$40.4\pm2.3$\\ $(48.9)$\end{tabular}} &
  \multicolumn{1}{c|}{\begin{tabular}[c]{@{}c@{}}$53.9\pm3.1$\\ $(65.8)$\end{tabular}} &
  \multicolumn{1}{c|}{\begin{tabular}[c]{@{}c@{}}-\\ $(72.9)$\end{tabular}} &
  \multicolumn{1}{c|}{\begin{tabular}[c]{@{}c@{}}-\\ $(83.0)$\end{tabular}} \\ \cline{2-7} 
\multicolumn{1}{|c|}{} &
  2000 &
  \multicolumn{1}{c|}{\begin{tabular}[c]{@{}c@{}}$24.9\pm0.5$\\ $(25.6)$\end{tabular}} &
  \multicolumn{1}{c|}{\begin{tabular}[c]{@{}c@{}}$29.2\pm0.1$\\ $(37.3)$\end{tabular}} &
  \multicolumn{1}{c|}{\begin{tabular}[c]{@{}c@{}}$35.3\pm0.4$\\ $(48.6)$\end{tabular}} &
  \multicolumn{1}{c|}{\begin{tabular}[c]{@{}c@{}}$46.5\pm1.4$\\ $(62.8)$\end{tabular}} &
  \multicolumn{1}{c|}{\begin{tabular}[c]{@{}c@{}}-\\ $(76.4)$\end{tabular}} \\ \cline{2-7} 
\multicolumn{1}{|c|}{} &
  4000 &
  \multicolumn{1}{c|}{\begin{tabular}[c]{@{}c@{}}$20.4\pm0.2$\\ $(18.6)$\end{tabular}} &
  \multicolumn{1}{c|}{\begin{tabular}[c]{@{}c@{}}$22.8\pm0.1$\\ $(24.9)$\end{tabular}} &
  \multicolumn{1}{c|}{\begin{tabular}[c]{@{}c@{}}$26.7\pm0.2$\\ $(37.2)$\end{tabular}} &
  \multicolumn{1}{c|}{\begin{tabular}[c]{@{}c@{}}$33.9\pm2.7$\\ $(47.7)$\end{tabular}} &
  \multicolumn{1}{c|}{\begin{tabular}[c]{@{}c@{}}$38.1\pm1.7$\\ $(60.6)$\end{tabular}} \\ \cline{2-7} 
\multicolumn{1}{|c|}{} &
  8000 &
  \multicolumn{1}{c|}{\begin{tabular}[c]{@{}c@{}}$17.2\pm1.2$\\ $(12.0)$\end{tabular}} &
  \multicolumn{1}{c|}{\begin{tabular}[c]{@{}c@{}}$17.6\pm0.2$\\ $(17.8)$\end{tabular}} &
  \multicolumn{1}{c|}{\begin{tabular}[c]{@{}c@{}}$22.0\pm0.1$\\ $(24.0)$\end{tabular}} &
  \multicolumn{1}{c|}{\begin{tabular}[c]{@{}c@{}}$24.1\pm0.3$\\ $(35.6)$\end{tabular}} &
  \multicolumn{1}{c|}{\begin{tabular}[c]{@{}c@{}}$27.2\pm0.7$\\ $(51.4)$\end{tabular}} \\ \hline
\end{tabular}
\caption{Critical latitudes as determined in Figure \ref{fig:CL_table}. For each simulation, the average critical latitude from both hemispheres is given along with an error estimate based on the difference between the two values. The number between brackets denotes the theoretical estimate of the critical latitude using equation (\ref{eq:CL}).}
\label{tab:CLtable}
\end{table}

\begin{figure}[hbt]
\centering
\includegraphics[width=.6\columnwidth]{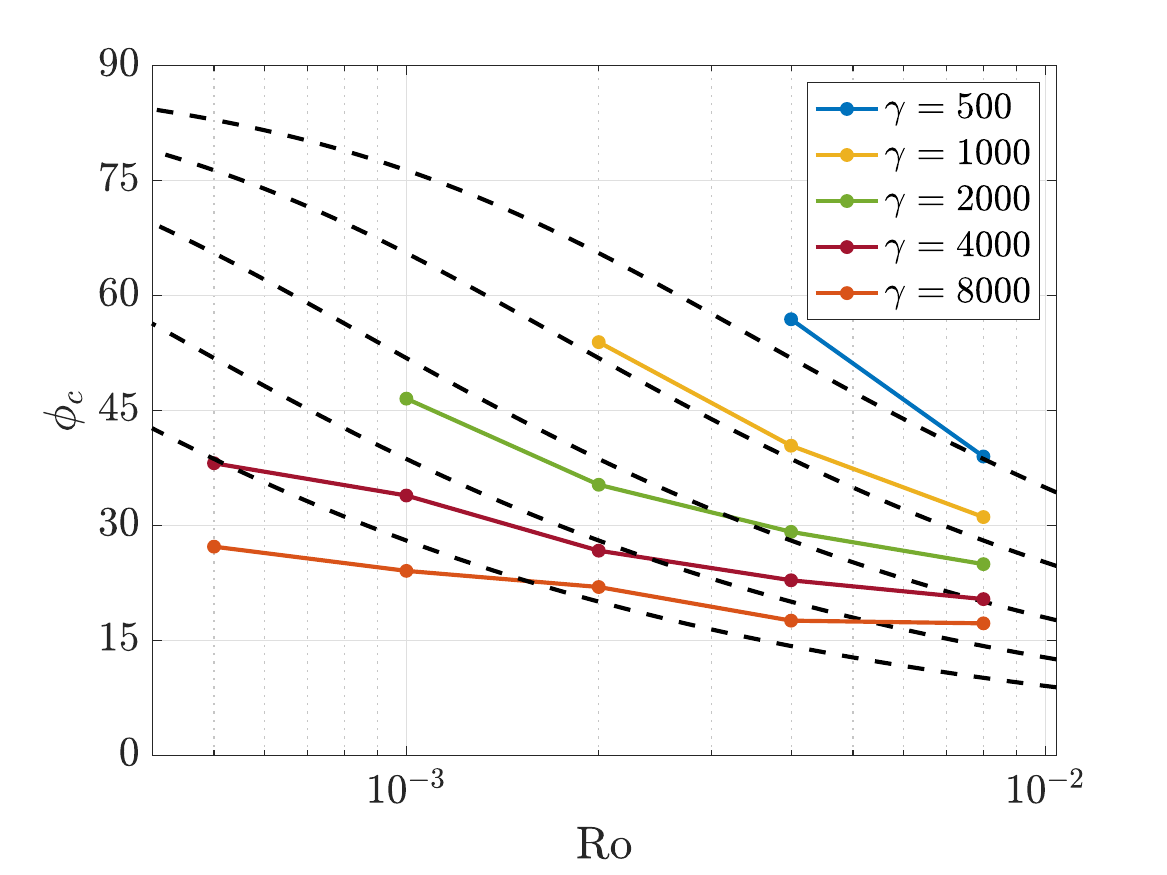}
\caption{Critical latitude $\phi_{cr}$ as a function of the Rossby number at different values of the Lamb parameter. The dashed lines represent the predicted critical latitude according to equation (\ref{eq:CL}). The reported critical latitudes of the simulations follow using the procedure described in figure \ref{fig:CL_table}. Simulations that show no critical latitude for jet formation are absent from this figure, which holds for cases where the ratio $\Omega/\gamma$ is high. Critical latitudes closely follow theoretical predictions at moderate values of $\gamma$ and $\Omega$. However, for simulations with a low predicted critical latitude, where the ratio $\Omega/\gamma$ is small, simulations show a much larger critical latitude.  }
\label{fig:critlat}
\end{figure}

The calculated values of the critical latitude are summarized in Table \ref{tab:CLtable}. For each simulation, the table shows an average of the critical latitude determined from both hemispheres, as well as the difference between these values which gives an error estimate for the calculated values. In brackets, the theoretical estimate of the critical latitude according to equation (\ref{eq:CL}) is given. These values are also visualized in Figure~\ref{fig:critlat}. This figure shows the critical latitude as a function of the Rossby number at the different values of the Lamb parameter. The results show that a separation between flow regimes is clearly visible in the regime of moderate Rossby and Lamb parameters. There, the simulation results also correspond qualitatively with the approximate theoretical values. Conversely, such separation of flow regimes does not exist in the cases of strong rotation and weak stratification. This further quantifies the well-established notion that the geostrophic balance introduces an equatorward transfer of kinetic energy where the flow develops into jets~\cite{theiss2004equatorward,vallis2017atmospheric}. In the cases where the Lamb parameter is small, this effect is very small, and an almost uniform emergence of jets is observed on the planet. In the limiting case where $\gamma=0$, the QG equation reverts to the rotating 2D Euler equations~\cite{luesink2021stochastic}, which features no attenuation of jet amplitudes at the poles at all~\cite{franken2024zeitlin}.

In the geostrophic regime, corresponding to moderate values of the Rossby number and Lamb parameter, the observed values of the critical latitude agree closely with their theoretical counterparts. This confirms the notion that the ability of jets to form is largely determined by the nonlinearity parameter (\ref{eq:nonlinparam}). In the cases where a low critical latitude is predicted (at large Lamb parameters), the theoretical estimates are systematically lower than those of simulations. This seems to be caused by the finite width of the jets, which makes it impossible for them to be further confined in their meridional extent. In the bottom row of Figure~\ref{fig:Vzon_table}, it appears that only one pair of jets can form in these cases, whose width determines the critical latitude. This limitation is absent from the theoretical considerations, and was avoided in \cite{theiss2004equatorward} by using a tangent plane domain extending beyond the equator.

To our knowledge, this is the first quantitative study on the value of the critical latitude in geostrophic turbulence. The insights gained from this study pave the way for future application of fully structure-preserving methods to global fluid dynamics. Investigations into the impact of more complex topographies and boundary conditions could lead a better understanding of the dynamics of zonal jet formation in Earth applications. Additionally, this model can be used to simulate specific planetary atmospheres, such as those of Jupiter and Saturn, to shed more light on the dynamics and stability of jets.

\section{Concluding remarks}
\label{sec:conclu}

This paper presented a detailed numerical investigation into the features of global geostrophic turbulence in the absence of forcing and dissipation. This type of simulation is enabled by the Casimir-preserving numerical method developed in~\cite{franken2024zeitlin}. The study focused on the development of large zonal jets, which are prominent in many geophysical flows and particularly noticeable in the weather layers of giant gas planets. These jets form due to the east-west stretching of vortices that coalesce into elongated structures and eventually into circumnavigating jets, a process that depends heavily on the gradient of the Coriolis parameter, leading to a critical latitude beyond which no jets form. 

The global quasi-geostrophic model enabled a detailed study of this critical latitude, considering the fully latitude-dependent Coriolis parameter. Based on theoretical estimations, the critical latitude was predicted to only depend on the product of the Rossby number and the Lamb parameter~\cite{theiss2004equatorward}. Through multiple simulations of initially random flow fields, the critical latitude was determined from the zonally averaged amplitudes of velocity profiles as a function of these parameters.

The results indicated a close alignment with geostrophic theory, especially near typical Rossby and Lamb parameter values for Earth's atmosphere, corroborating earlier findings by \cite{theiss2004equatorward} and \cite{klocker2016regime} on a rectangular tangent plane. However, in regimes of weak rotation (high Rossby numbers) and strong stratification (high Lamb parameter values), a clear critical latitude did not emerge. Instead, the amplitude and width of zonal jets gradually decreased toward the poles of the sphere. This research opens avenues for further theoretical studies on jet formation under the influence of a fully latitude-dependent Coriolis parameter.

\section*{Acknowledgement}

The authors gratefully acknowledge the valuable discussions with Darryl Holm (Imperial College London) and Paolo Cifani. This publication is part of the project SPRESTO which is financed by the Dutch Research Council (NWO) in their TOP1 program. Simulations were carried out on the Dutch national e‐infrastructure with the support of SURF Cooperative.

\bibliographystyle{plain}
\bibliography{references}

\end{document}